\documentclass[twocolumn,showpacs,preprintnumbers,amsmath,amssymb,prb]{revtex4}

\usepackage{graphicx}
\usepackage{dcolumn}
\usepackage{bm}

\begin{document}


\title{Single-electron tunneling in InP nanowires}

\author{S. De Franceschi$^1$, J. A. van Dam$^1$, E. P. A. M. Bakkers$^2$,
L. F. Feiner$^2$, L. Gurevich$^1$, and L. P. Kouwenhoven$^1$}
\affiliation{{$^1$Department of NanoScience and ERATO Mesoscopic Correlation Project,Delft University of Technology,
PO Box 5046, 2600 GA Delft, The Netherlands} \\
{$^2$Philips Research Laboratories, Prof. Holstlaan 4, 5656 AA Eindhoven, The Netherlands}
}

\date{\today}

\begin{abstract}

We report on the fabrication and electrical characterization of field-effect devices based on wire-shaped InP crystals grown from Au catalyst particles by a vapor-liquid-solid process. Our InP wires are n-type doped with diameters in the 40$-$55 nm range and lengths of several $\mu$m. After being deposited on an oxidized Si substrate, wires are contacted individually via e-beam fabricated Ti/Al electrodes. We obtain contact resistances as low as $\sim$10 k$\Omega$, with minor temperature dependence. The distance between the electrodes varies between 0.2 and 2 $\mu$m. The electron density in the wires is changed with a back gate.
Low-temperature transport measurements show Coulomb-blockade behavior with
single-electron charging energies of $\sim$1 meV. We also demonstrate energy quantization resulting from the confinement in the wire.

\end{abstract}


\maketitle

Chemically synthesized semiconductor nanowires (or nanowhiskers) attract increasing interest as building blocks for a bottom-up approach to the fabrication of nanoscale devices and sensors. A key property of these material systems is the unique versatility in terms of geometrical dimensions and composition. Nanowires have already been grown from several semiconductor materials (group-IV elements \cite{Cui,Wu}, III-V \cite{Hiruma,Duan,Bjork1,Huang2} and II-VI compounds \cite{Solanki}), including structures with variable doping and composition, such as n-type/p-type InP \cite{Gudiksen}, InAs/InP \cite{Bjork1}, GaAs/GaP \cite{Gudiksen}, and Si/SiGe \cite{Wu}.
The growth technique is based on the vapor-liquid-solid (VLS) process \cite{Wagner} occurring at metallic catalysts, such as nanometer-sized Au particles. The nanowire diameter is set by the catalyst dimension, typically
10$-$100 nm. The nanowire length is proportional to the growth time and can exceed hundreds of microns.
The semiconductor is provided either by metalorganic vapor-phase sources \cite{Hiruma, Bjork1}, or by laser ablation \cite{Morales}. Many room-temperature applications have already been shown, such as single-nanowire field-effect transistors (FETs) \cite{Duan}, diodes \cite{Gudiksen}, and logic gates \cite{Huang} combining both n-type and p-type nanowires. Very recently, nanowire heterostructures have been operated as resonant tunneling diodes
at 4.2 K \cite{Bjork2}. Yet the low-temperature properties of nanowires and their potential for novel quantum devices are still widely unexplored.

In this Letter we describe the realization of FET devices from individual n-type InP nanowires. We discuss their transport properties down to 0.35 K, where single-electron tunneling and quantum effects can play a dominant role.
We also provide details on the fabrication of the electrical contacts to the nanowires, a crucial aspect of the present work.

Our InP nanowires are grown via the laser-assisted VLS method. A pulsed laser (193-nm ArF laser, 10 Hz, 100 mJ/pulse) is used to ablate from a pressed InP powder enriched with 1 mol\% of Se, which acts as a donor impurity in InP. Nanowires grow from Au seeds formed after annealing a 2 \AA-equivalent Au film deposited on a Si substrate with a superficial
native oxide. During growth time ($\sim$30 min) the substrate temperature is
kept at 475 $^\circ$C. The resulting nanowires are
5$-$10 $\mu$m long with a diameter of 40$-$55 nm.

\begin{figure}
\includegraphics{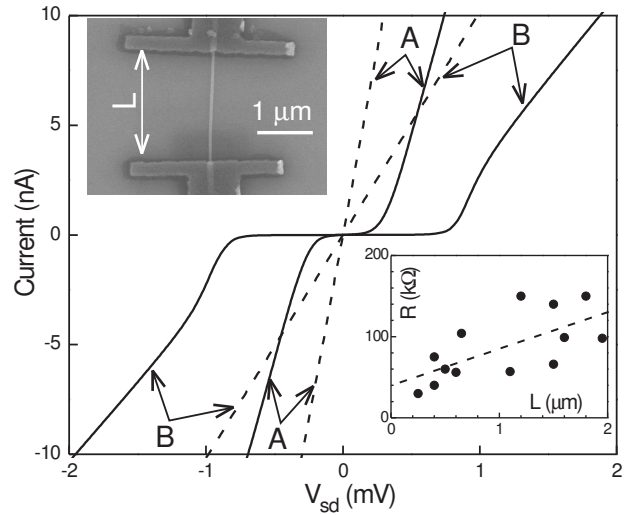}
\caption{Current-voltage characteristics at room temperature (dashed lines)
and 0.35 K (solid lines), for devices A and B.
Upper inset: scanning-electron micrograph of device B.
Lower inset: length dependence of the room-temperature source-drain resistance. Each solid circle refers to a different device. The dashed line is a linear fit.}
\end{figure}

Immediately after growth, the nanowires are dispersed in Chlorobenzene. A few droplets of this dispersion are deposited on a p$^{+}$ Si substrate with a 250-nm-thick SiO$_2$ overlayer. To favor the adhesion of the nanowires, the surface is functionalized with a self-assembled monolayer
of 3-aminopropyltriethoxysilane (APTES) \cite{Liu}.
Optical imaging is used to locate the nanowires with respect to a reference pattern of predefined Pt markers. The nanowires are then individually contacted with a pair of metal electrodes (source and drain leads) defined by electron-beam lithography (see upper inset to Fig. 1). The distance, $L$, between the source and drain electrodes is varied between 0.2 and 2 $\mu$m.

The contact electrodes consist of thermally evaporated Ti(100 nm)/Al(20 nm). Before metal deposition, samples are treated with BHF for 20 s in order to etch the oxide layer around the nanowires \cite{footnote2}. As-deposited contacts show high resistance, typically in excess of 10 G$\Omega$. The contact resistance improves drastically after forming-gas rapid-thermal annealing at 475 $^\circ$C for 60 s (for a discussion of the interface reaction between Ti and InP we refer to Ref. 15).

We have characterized over ten devices at different temperatures, $T$. At room temperature, current-voltage ($I-V$) characteristics are linear (see dashed lines in Fig. 1) with resistances, $R$, as low as 30 k$\Omega$. Despite sample-to-sample fluctuations, $R$ appears to increase with $L$, as shown in the lower inset to Fig. 1. A linear fit yields $R = 40$ k$\Omega + 45$ k$\Omega/\mu$m$ \times L$, where the constant term and the slope coefficient can be taken as rough estimates of the total contact resistance and the wire resistivity, respectively.

Below a few Kelvin, the $I-V$ characteristics develop a non-linearity around zero bias. This behavior, common to all measured devices, is shown in Fig. 1 for device $A$ ($L=0.20$ $\mu$m), and $B$ ($L= 1.95$ $\mu$m), respectively the shortest and longest devices measured. The zero-bias suppression of the conductance has a pronounced dependence on the voltage, $V_g$, applied to the p$^+$ Si substrate. This is a characteristic fingerprint of Coulomb-blockaded transport which becomes dominant at low temperatures.  In fact, an electronic island, formed inside the nanowire segment between source and drain electrodes, leads to Coulomb blockade of transport when $k_B T < e^2/C$, where $C$ is the total capacitance of the island \cite{CB}.
At source-drain voltages, $V_{sd}$, larger than $e/C$, the slope of the low-$T$ trace is close to the corresponding room-$T$ value, indicating little $T$-dependence of the contact resistance.

Figure 2a shows conductance, $G$, versus $V_g$ for device C ($L= 0.65$ $\mu$m) and D ($L= 1.6$ $\mu$m, inset). Both traces exhibit sharp peaks corresponding to Coulomb-blockade oscillations. This clearly demonstrates that we have achieved single-electron control over the electronic charge and the transport properties of the nanowire. The Coulomb peaks have irregularly distributed sizes, and their $V_g$-spacing varies considerably, suggesting the formation of more than one electronic island along the nanowire.
This interpretation is supported by the measurement shown in Fig. 2b where the differential conductance, $dI/dV_{sd}$, of device C is plotted on gray scale as a function of ($V_g$, $V_{sd}$). In this plot, Coulomb blockade takes place within dark regions with the characteristic diamond shape.  In some cases, such as for  $V_g$ between $-$40 and $-$90 mV, Coulomb diamonds are clearly separated from each other and have all their edges fully defined. This is characteristic of Coulomb-blockaded transport through a single electronic island. In other $V_g$-regions, however, diamonds overlap with each other, as we would expect for a nanowire containing more than one (most likely two)
islands in series.
The $V_g$-dependent alternation of single- and double-island regimes, shown in Fig. 2b, is representative of the general behavior in our devices. We would like to stress that such charge reconfigurations are found to be very stable and reproducible.

\begin{figure}
\begin{center}
\includegraphics{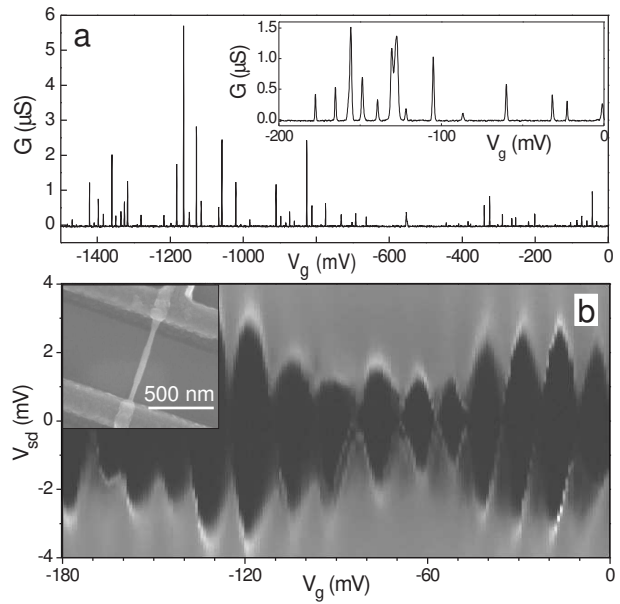}
\caption{(a) Conductance, $G$, versus back-gate voltage, $V_g$, measured at 0.35 K with a dc bias $V_{sd}= 20 \mu$V. The two traces refer to device C, and D (inset). (b) Gray-scale plot of differential conductance, $dI/dV_{sd}$,
versus ($V_g$, $V_{sd}$). $dI/dV_{sd}$ increases when going from dark to light
gray. The measurement refers to device C, and was taken at 0.35 K
with lock-in technique at an ac bias excitation of 20 $\mu$V.
Inset: scanning-electron micrograph of device C.}
\end{center}
\end{figure}

Each Coulomb diamond is associated with a well-defined number of confined electrons, $N$. From the half-height (along $V_{sd}$) of the diamonds we estimate a charging energy $e^2/C \sim 1$ meV. The $V_g$-width of the diamonds is around 10$-$20 mV, from which we deduce $C_g/C \approx 1/7$, where  $C_g$ is the capacitance to the back gate \cite{footnote3}.
This implies that $N$ decreases by $\sim$100 when moving from right to left in Fig. 2a. Based on separate studies \cite{footnote1}, we believe this is still only a small fraction of the total amount of conduction electrons in the nanowire.

We now focus on a small $V_g$-range where device C exhibits single-island behavior.
Figure 3 shows several $dI/dV_{sd}$$-$vs$-$$V_g$ traces taken at different values of $V_{sd}$ between 0 and $-$0.6 mV. The lowest trace ($V_{sd}=0$) shows two Coulomb peaks denoting transitions between successive charge states: say from $N-1$ to $N$ (left peak), and from $N$ to $N+1$ (right peak).
At finite bias, each peak in $dI/dV_{sd}$ splits proportionally to $V_{sd}$, as expected from ordinary Coulomb-blockade theory. The left-moving (right-moving) split-peak corresponds to the onset of tunneling from (to) the source (drain) lead. Interestingly, at larger $V_{sd}$ extra resonances appear between the split-peaks. Increasing $V_{sd}$, the $V_{g}$-positions of such resonances evolve parallel to one of the split-peaks, as emphasized by dashed lines. We recognize this behavior as characteristic of transport through a quantum-dot system with a discrete energy spectrum \cite{Johnson}. The two extra resonances can be readily explained as the result of tunneling processes involving excited states of the quantum dot.
The resonance on the right side of Fig. 3 denotes the onset of tunneling from the source lead to the first excited state with $N+1$ electrons (see top-right inset). The one on the left can be ascribed to the tunneling of an electron from the dot to the drain which leaves the dot in an excited state with $N-1$ electrons (see top-left inset).
The bias voltage for which the resonance associated to an $N$-electron excited-state begins to emerge is a direct measurement of the corresponding excitation energy, $\Delta E(N)$. We find $\Delta E (N-1) \approx 0.36$ meV and $\Delta E(N+1)  \approx 0.30$ meV.

\begin{figure}
\begin{center}
\includegraphics{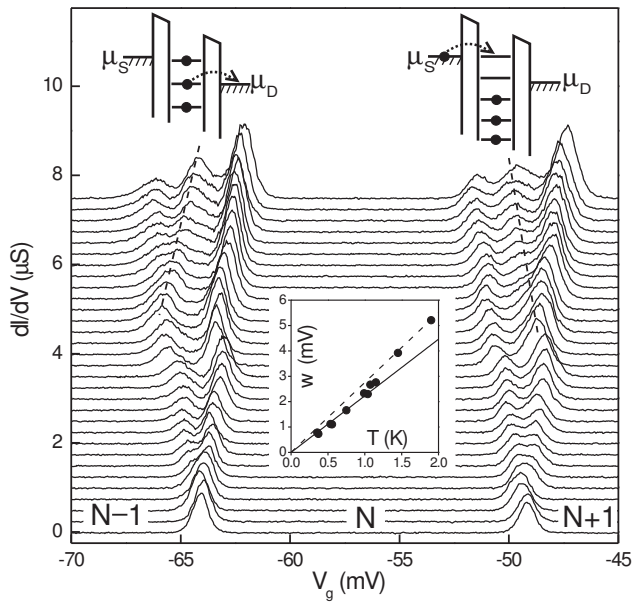}
\caption{$dI/dV_{sd}$ vs $V_g$, for different dc values of $V_{sd}$,
from 0 (lower trace) to $-$0.6 mV (upper trace) in steps of $-$0.02 mV.
Dashed lines indicate the evolution of the peaks associated with tunneling
via excited states. A simplified picture of the corresponding processes is given in the top insets. Bottom inset: full-width at half maximum, $w$, vs temperature, $T$, for the left Coulomb peak at $V_{sd}=0$.
The solid (dashed) line is the theoretical prediction,
$w = (C/eC_g) \times 3.52 k_B T$ ($w = (C/eC_g) \times 4.35 k_B T$), for the quantum (classical) regime \cite{Foxman}. We used $C/C_g= 7.35$, obtained from the
$V_{sd}$-dependence of the $V_g$-splitting.}
\end{center}
\end{figure}

Another sign of energy quantization is based on the $T$-dependence of the conductance peaks at $V_{sd} = 0$.
As expected for single-electron tunneling in quantum dots, lowering $T$ gives an increased peak height as a result of resonant tunneling \cite{CB}.
This is in fact observed in the low-$T$ limit (data not shown). The full-width at half maximum increases linearly with $T$, but with two different slopes, associated with the quantum ($k_B T < \Delta E$) and the classical ($k_B T > \Delta E$) regime \cite{Foxman}. In our case, the transition between the two regimes occurs at about 1.2 K, (see Fig. 3b) corresponding to $\Delta E \sim 0.1$ meV, in agreement with the previous findings.
Our observation of a discrete energy spectrum represents an important premise for a deeper investigation of quantum phenomena and the development of controllable quantum devices based on semiconductor nanowires.

We thank T. Nolst Trenit\'e and S. Tarucha for help and discussions.
We acknowledge financial support from the Specially Promoted Research, Grant-in-Aid for Scientific Research, from the Ministry of Education, Science and Culture in Japan, and from the Dutch Organisation for Fundamental Research on Matter (FOM).



\newpage


\begin{thebibliography}{99}
\bibitem{footnote3}
From a simplified model we estimate $C_g \approx [2 \pi \epsilon \epsilon_0$/ln$(2d/R)] \times L_0$, where $L_0$ is the length of a cylindrical electronic island, $\epsilon=4$ (relative dielectric constant for Si0$_2$), $R= 20$ nm (island radius), $d = 250$ nm (distance to the back gate). Then $U = e^2/C = 1$ meV and $C_g/C=1/7$ yield $L_0 \approx 0.3$ $\mu$m $\lesssim L$.
\bibitem{footnote1} We studied the pinch-off characteristic of similar devices, fabricated from the same batch of nanowires. Using a simple electrostatics model (see,
e.g., Ref. 6), we estimated a linear density of $\sim$$10^4$ electrons/$\mu$m. Neglecting surface-induced depletion, this corresponds to $\sim$10$^{19}$ electrons/cm$^3$. Secondary-ion mass-spectroscopy measurements, performed on a massive amount of
nanowires grown under the same conditions, gave a Se content of $10^{19} - 10^{20}$ atoms/cm$^3$.
\end{thebibliography}
\end{document}